\documentclass[prb,twocolumn,superscriptaddress,showpacs,showkeys]{revtex4-1}
\usepackage{graphicx}
\usepackage{amssymb}
\usepackage{multirow}

\newcommand{\ped}[1]{\ensuremath{_{\rm #1}}}
\newcommand{\apex}[1]{\ensuremath{^{\rm #1}}}
\begin{document}
\title{Proximity Eliashberg theory of electrostatic field-effect-doping in superconducting films}
\author{G. A. Ummarino}
\email{giovanni.ummarino@polito.it}
\affiliation{Istituto di Ingegneria e Fisica dei Materiali, Dipartimento di Scienza Applicata e Tecnologia, Politecnico di Torino, Corso Duca degli Abruzzi 24, 10129 Torino, Italy}
\affiliation{National Research Nuclear University MEPhI (Moscow Engineering Physics Institute), Kashira Hwy 31, Moskva 115409, Russia}
\author{E. Piatti}
\author{D. Daghero}
\author{R. S. Gonnelli}
\affiliation{Istituto di Ingegneria e Fisica dei Materiali, Dipartimento di Scienza Applicata e Tecnologia, Politecnico di Torino, Corso Duca degli Abruzzi 24, 10129 Torino, Italy}
\author{Irina Yu. Sklyadneva}
\affiliation{Donostia International Physics Center (DIPC), 20018 San Sebastian/Donostia, Basque Country, Spain}
%\
\author{E. V. Chulkov}
\affiliation{Donostia International Physics Center (DIPC), 20018 San Sebastian/Donostia, Basque Country, Spain}
\affiliation{St. Petersburg State University, 199034, St. Petersburg, Russian Federation}
\affiliation{Departamento de Fisica de Materiales, Facultad de Ciencias Quimicas, UPV/EHU, Apdo. 1072, 20080 San Sebastian/Donostia, Basque Country, Spain}
\affiliation{Centro de Fisica de Materiales CFM - Materials Physics Center MPC, Centro Mixto CSIC-UPV/EHU, 20018 San Sebastian/Donostia, Basque Country, Spain.}
\author{R. Heid}
\affiliation{Karlsruher Institut f\"{u}r Technologie, Institut f\"{u}r Festk\"{o}rperphysik, D-76021 Karlsruhe, Germany}
\begin{abstract}
We calculate the effect of a static electric field on the critical temperature of a s-wave one band superconductor in the framework of proximity effect Eliashberg theory.
In the weak electrostatic field limit the theory has no free parameters while, in general, the only free parameter is the thickness of the surface layer where the electric field acts.
We conclude that the best situation for increasing the critical temperature is to have a very thin film of a superconducting material with a strong increase of electron-phonon (boson) constant upon charging.
\end{abstract}
\maketitle
\section{INTRODUCTION}
In the last decade, electric double layer (EDL) gating has come to the forefront of solid state physics due to its capability to tune the surface carrier density of a wide range of different materials well beyond the limits imposed by solid-gate field-effect devices. The order-of-magnitude enhancement in the gate electric field allows this technique to reach doping levels comparable to those of standard chemical doping. This is possible due to the extremely large specific capacitance of the EDL that builds up at the interface between the electrolyte and the material under study \cite{FujimotoReview2013,UenoReview2014,GoldmanReview2014,SaitoReview2016}.

EDL gating was first exploited to control the surface electronic properties of relatively low-carrier density systems, where the electric field effect is more readily observable. Field-induced superconductivity was first demonstrated in strontium titanate \cite{UenoNatureMater2008} and zirconium nitride chloride \cite{YeNatureMater2010,SaitoScience2015}, and subsequently on other insulating systems such as perovskites \cite{UenoNatureNano2011} and layered transition-metal dichalcogenides \cite{YeScience2012,JoNanoLett2015,LuScience2015,ShiSciRep2015,YuNatNano2015,CostanzoNatNano2016,SaitoNatPhys2016}. Significant effort was also invested in the control of the superconducting properties of cuprates \cite{bollinger11,LengPRL2011,LengPRL2012,MaruyamaAPL2015,JinSciRep2016,FeteAPL2016,BurlachkovPRB1993,GhinovkerPRB1995,walter16}, although in this class of materials the mechanism behind the carrier density modulation is still debated \cite{walter16}.

More recently however, several experimental studies have been devoted to the exploration of field effect in superconductors \cite{libro} with a large ($\gtrsim 1 \cdot 10^{22}$ cm$\apex{-3}$) native carrier density. The interplay between two different ground states, namely superconductivity and charge density waves, was explored in titanium and niobium diselenides \cite{LiNature2016,YoshidaAPL2016,XiPRL2016}. The thickness and gate voltage dependence of a high-temperature superconducting phase were studied in iron selenide, both in thin-film \cite{ShiogaiNaturePhys2015} and thin flake \cite{LeiPRL2016,HanzawaPNAS2016} forms. The effect of the ultrahigh interface electric fields achievable via EDL gating were also probed in standard BCS superconductors, namely niobium \cite{ChoiAPL2014} and niobium nitride \cite{PiattiJSNM2016,PiattiNbN}.

With the exception of the work of Ref.\onlinecite{XiPRL2016} on niobium diselenide, all of these studies have been performed on relatively thick samples ($\gtrsim 10$ nm) with a thickness larger than the electrostatic screening length. These systems are thus expected to develop a strong dependence of their electronic properties along the $z$ direction (z being perpendicular to the sample surface). As a first approximation, this dependence can be conceptualized by schematizing the system as the parallel of a surface layer (where the carrier density is modified by the electric field) and an unperturbed bulk. The two electronic systems can be expected to couple via superconducting proximity effect, resulting in a complicated response to the applied electric field that goes well beyond a simple modification of the superconducting properties of the surface layer alone \cite{PiattiNbN} and is strongly dependent on both the electrostatic screening length and the total thickness of the film.

So far, the only quantitative assessment of this phenomenon has been reported in the framework of the strong-coupling limit of the BCS theory of superconductivity \cite{PiattiNbN}. A proper theoretical treatment for field effect on more complex materials, which can be described only by means of the more complete Eliashberg theory \cite{carbibastardo,ummarinorev}, is lacking. Given the rising interest in the control of the properties of superconductors by means of surface electric fields, the development of such a theoretical approach would be very convenient not only to quantitatively describe the results of future experiments, but also to determine beforehand the experimental conditions (e.g., device thickness) most suitable for an optimal control of the superconducting order via electric fields.

\begin{figure}
\begin{center}
\includegraphics[keepaspectratio, width=\columnwidth]{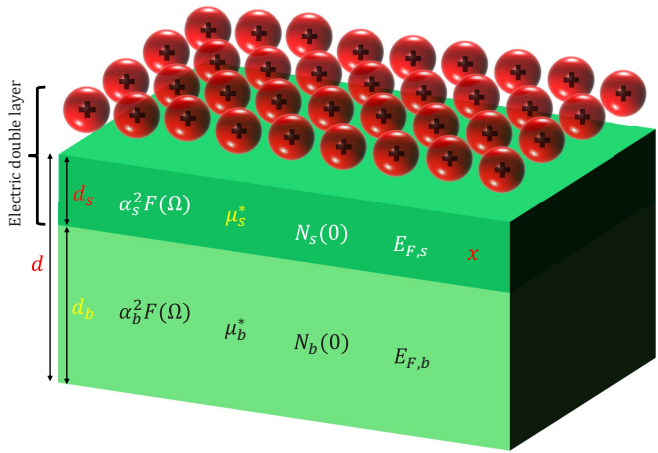}
\vspace{-5mm} \caption{(Color online)
Scheme of an EDL-gated superconducting thin film. The layer of adsorbed ions and the surface layer where the carrier density is perturbed (dark green region) compose the EDL. The unperturbed bulk of the film is indicated in light green color. For both layers, we indicate the relevant parameters of the proximity Eliashberg equations (see text for details). Parameters in red, black and white indicate the free parameters of the model, data obtained from the literature, and the output of the DFT calculations respectively. Parameters in yellow are obtained from these by simple calculations.}\label{Figure0}
\end{center}
\end{figure}

In this work, we use the Eliashberg theory of proximity effect to describe a junction composed by the perturbed surface layer ($T_{c}=T_{c,s}$), where the carrier density is modulated (with a doping level per unitary cell $x$), and the underlying unperturbed bulk ($T_{c}=T_{c,b}$). Here $s$ and $b$ indicate ``surface" and ``bulk" respectively (see Fig. \ref{Figure0}). Under the application of an electric field, $T_{c,s}\neq T_{c,b}$ and the material behaves like a junction between a superconductor and a normal metal in the temperature range bounded by $T_{c,s}$ and $T_{c,b}$. If the application of the electric field increases (decreases) $T_{c,s}$, then the surface layer will be the superconductor (normal metal) and the bulk will be the normal metal (superconductor). We perform the calculation for lead since all the input parameters of the theory are well-known in the literature for this strong-coupling superconductor \cite{carbibastardo}.

%%%%%%%%%%%%%%%%%%%%%%%%%%%%%%%%%%%%%%%%%%%%%%%%%%%%%%%%%%%%%%%%%%
\section{MODEL: PROXIMITY ELIASHBERG EQUATIONS}
In general, a proximity effect at a superconductor/normal metal junction is observed as the opening of a finite superconducting gap in the normal metal together with its reduction in a thin region of the superconductor close to the junction.
In our model we use the one band s-wave Eliashberg equations \cite{carbibastardo,ummarinorev}
with proximity effect to calculate the critical temperature of the system.
In this case we have to solve four coupled equations for the gaps $\Delta_{s,b}(i\omega_{n})$
and the renormalization functions $Z_{s,b}(i\omega_{n})$, where $\omega_{n}$ are the Matsubara frequencies. The imaginary-axis
equations with proximity effect \cite{Mc,Carbi1,Carbi2,Carbi3,kresin} are:
\begin{eqnarray}
&&\omega_{n}Z_{b}(i\omega_{n})=\omega_{n}+ \pi T\sum_{m}\Lambda^{Z}_{b}(i\omega_{n},i\omega_{m})N^{Z}_{b}(i\omega_{m})+\nonumber\\
&&+\Gamma\ped{b} N^{Z}_{s}(i\omega_{n})
\label{eq:EE1}
\end{eqnarray}
\begin{eqnarray}
&&Z_{b}(i\omega_{n})\Delta_{b}(i\omega_{n})=\pi
T\sum_{m}\big[\Lambda^{\Delta}_{b}(i\omega_{n},i\omega_{m})-\mu^{*}_{b}(\omega_{c})\big]\times\nonumber\\
&&\times\Theta(\omega_{c}-|\omega_{m}|)N^{\Delta}_{b}(i\omega_{m})
+\Gamma\ped{b} N^{\Delta}_{s}(i\omega_{n})\phantom{aaaaaa}
 \label{eq:EE2}
\end{eqnarray}
\begin{eqnarray}
&&\omega_{n}Z_{s}(i\omega_{n})=\omega_{n}+ \pi T\sum_{m}\Lambda^{Z}_{s}(i\omega_{n},i\omega_{m})N^{Z}_{s}(i\omega_{m})+\nonumber\\
&&\Gamma\ped{s} N^{Z}_{b}(i\omega_{n})
\label{eq:EE3}
\end{eqnarray}
\begin{eqnarray}
&&Z_{s}(i\omega_{n})\Delta_{s}(i\omega_{n})=\pi
T\sum_{m}\big[\Lambda^{\Delta}_{s}(i\omega_{n},i\omega_{m})-\mu^{*}_{s}(\omega_{c})\big]\times\nonumber\\
&&\times\Theta(\omega_{c}-|\omega_{m}|)N^{\Delta}_{s}(i\omega_{m})
+\Gamma\ped{s}N^{\Delta}_{b}(i\omega_{n})\phantom{aaaaaa}
 \label{eq:EE4}
\end{eqnarray}
where $\mu^{*}_{s(b)}$ are the Coulomb pseudopotentials in the surface and in the bulk respectively, $\Theta$ is the Heaviside function and $\omega_{c}$ is a cutoff
energy at least three times larger than the maximum phonon energy. Thus we have
\begin{equation}
\Lambda_{s(b)}(i\omega_{n},i\omega_{m})=2
\int_{0}^{+\infty}d\Omega \Omega
\alpha^{2}_{s(b)}F(\Omega)/[(\omega_{n}-\omega_{m})^{2}+\Omega^{2}]
\end{equation}
\begin{equation}
\Gamma_{s(b)}=\pi|t|^{2}Ad_{b(s)}N_{b(s)}(0)
\label{eq:EE6}
\end{equation}
and thus $\frac{\Gamma_{s}}{\Gamma_{b}}=\frac{d_{b}N_{b}(0)}{d_{s}N_{s}(0)}$,
\begin{equation}
N^{\Delta}_{s(b)}(i\omega_{m})=\Delta_{s(b)}(i\omega_{m})/
{\sqrt{\omega^{2}_{m}+\Delta^{2}_{s(b)}(i\omega_{m})}}
\end{equation}
and
\begin{equation}
N^{Z}_{s(b)}(i\omega_{m})=\omega_{m}/{\sqrt{\omega^{2}_{m}+\Delta^{2}_{s(b)}(i\omega_{m})}}
\end{equation}
where $\alpha^{2}_{s(b)}F(\Omega)$ are the electron-phonon spectral functions, $A$ is the junction cross-sectional area, $|t|^{2}$ the transmission matrix equal to one in our case because the material is the same, $d_{s}$ and $d_{b}$ are the surface and bulk layer thicknesses respectively, such that ($d_{s}+d_{b}=d$ where $d$ is the total film thickness) and $N_{s(b)}(0)$ are the densities of states at the Fermi level $E_{F,s(b)}$ for the surface and bulk material.
The electron-phonon coupling constants are defined as
\begin{equation}
\lambda_{s(b)}=2\int_{0}^{+\infty}d\Omega\frac{\alpha^{2}_{s(b)}F(\Omega)}{\Omega}\nonumber\\
\end{equation}
and the representative energies as
\begin{equation}
ln(\omega_{ln,s(b)})=\frac{2}{\lambda_{s(b)}}\int_{0}^{+\infty}d\Omega ln\Omega \frac{\alpha^{2}_{s(b)}F(\Omega)}{\Omega}\nonumber\\
\end{equation}
The solution of these equations requires eleven
input parameters: the two electron-phonon spectral fuctions $\alpha^{2}_{s(b)}F(\Omega)$, the two Coulomb pseudopotentials $\mu^{*}_{s(b)}$, the values of the normal density of states at the Fermi level $N_{s(b)}(0)$, the shift of the Fermi energy $\Delta E_{F}=E_{F,s}-E_{F,b}$ that enters in the calculation of the surface Coulomb pseudopotential (as shown later), the value of the surface layer $d_{s}$, the film thickness $d$ and the junction cross-sectional area $A$.
The values of $d$ and $A$ are experimental data. The exact value of $d_{s}$, in particular in the case of very strong electric fields at the surface of a thin film, is in general difficult to determine \emph{a priori} \cite{PiattiNbN}. Thus, we leave it as a free parameter of the model, and we perform our calculations for different reasonable estimations of its value.

Typically, the bulk values of $\alpha^{2}_{b}F(\Omega)$, $\mu^{*}_{b}$, $N_{b}(0)$ and $E_{F,b}$ are known and can be found in the literature.
Thus, we assume that we need to determine only their surface values.
In the next Section, we will use density functional theory (DFT) to calculate $\alpha^{2}_{s}F(\Omega)$, $\Delta E_{F}$ and $N_{s}(0)$.

The value of the Coulomb pseudopotential in the surface layer $\mu^{*}_{s}$ can be obtained in the following way:
in the Thomas-Fermi theory where the dielectric function is \cite{UmmaC60}
$\varepsilon(q)=1+\frac{k^{2}_{TF}}{q^{2}}$
and the bare Coulomb pseudopotential $\mu$ is the angular average of the screened electrostatic potential
\begin{equation}
\mu=\frac{1}{4\pi^{2}\hbar v_{F}}\int _{0}^{2k_{F}}\frac{V(q)}{\varepsilon(q)}qdq
\label{equation_mu}
\end{equation}
Since
$V(q)=\frac{4 \pi e^{2}}{q^{2}}$
\cite{Morel} it turns out that

\begin{equation}
\mu=\frac{k^{2}_{TF}}{8 k^{2}_{F}}ln(1+\frac{4 k^{2}_{F}}{ k^{2}_{TF}}).
\end{equation}
Hence we write
\begin{equation}
\mu_{b}=\frac{a^{2}_{b}}{2}ln(1+\frac{1}{a^{2}_{b}}).
\end{equation}
with $a_{b}=2k_{TF,b}/k_{F,b}$.
Since $a_{b}$ can be calculated by numerically solving Eq. 13, and by remembering that the square of Thomas-Fermi wave number $k_{TF,b(s)}$ is proportional to $N_{b(s)}(0)$, we have
\begin{equation}
a^{2}_{s}=a^{2}_{b}(\frac{N_{s}(0)}{N_{b}(0)})/(\frac{1+\Delta E_{F}}{E_{F,b}})
\end{equation}
and thus
\begin{equation}
\mu_{s}=\frac{a^{2}_{s}}{2}ln(1+\frac{1}{a^{2}_{s}}).
\end{equation}
The new Coulomb pseudopotential \cite{Morel} in the surface layer is thus
\begin{equation}
\mu^{*}_{s}(\omega_{c})=\frac{\mu_{s}}{1+\mu_{s} ln((E_{F,b}+\Delta E_{F})/\omega_{c})}
\end{equation}
We note that, usually, the effect of electrostatic doping on $\mu^{*}$ is very small and can be neglected.
We can quantify the effect on $T_{c}$ of this small modulation of $\mu^{*}$ by computing it in the case of maximum doping $x=0.40$ and very thin film ($d=5$ nm), i.e. when the effect is largest.
As discussed in the next Section, the unperturbed Coulomb pseudopotential is $\mu^{*}(x=0)=0.14164$, while for the maximum doping Eqs. 12-16 give $\mu^{*}(x=0.4)=0.14048$.
If we use $d_{s}=d_{TF}$ we find respectively $T_{c}=7.3770$ K for the bulk (unperturbed) value of the Coulomb pseudopotential and $T_{c}=7.3768$ K for the surface value of the Coulomb pseudopotential.
Thus, if we consider the Coulomb pseudopotential to be doping-independent we underestimate the critical temperature of a $\Delta T_{c}|_{\Delta\mu^{*}}=-0.0002$ K ($\Delta T_{c}|_{\Delta\mu^{*}}/T_{c}=0.0027$ percent).

However, a possible critical situation can appear when the applied electric field is very strong and the Thomas-Fermi approximation does not hold anymore.
In such a case, $\mu^{*}$ becomes ill-defined as the Thomas-Fermi dielectric function is no longer strictly valid for very large electric fields.
Nevertheless, the true dielectric function $\varepsilon(q)$ should still be a function of the ratio $k_{TF}/k_{F}$ \cite{pastore}, which in the free-electron model is independent on the normal density of states at the Fermi level. Thus, Eq. \ref{equation_mu} should still be able to describe the behavior of the system as a first approximation.

\section{CALCULATION OF $\alpha^{2}_{s}F(\Omega)$, $\Delta E_{F}$ and $N_{s}(0)$}

DFT calculations are performed within the mixed-basis pseudopotential
method (MBPP) \cite{meyer}. For lead a norm-conserving relativistic
pseudopotential including $5d$ semicore states and partial core
corrections is constructed following the prescription of Vanderbilt
\cite{vanderbilt}. This provides both scalar-relativistic and spin-orbit
components of the pseudopotential.  Spin-orbit coupling (SOC) is
then taken into account within each DFT self-consistency cycle (for
more details on the SOC implementation see \cite{heid10}). The MBPP approach
utilizes a combination of local functions and plane waves for the basis set
expansion of the valence states, which reduces the size of the basis set
significantly.  One $d$ type local function is added at each lead atomic site
to efficiently treat the deep $5d$ potential. Sufficient convergence
is then achieved with a cutoff energy of 20 Ry for the plane waves.
The exchange-correlation it treated in the local density approximation
(LDA) \cite{hedin}. Brillouin zone (BZ) integrations are performed on regular
k-point meshes in conjuction with a Gaussian broadening of $0.2$ eV.
For phonons, $16 \times 16 \times 16$ meshes are used, while for the calculations of
density of states and electron-phonon coupling (EPC) even denser $32 \times 32 \times 32$
meshes are employed.

Phonon properties are calculated with the density-functional perturbation
theory \cite{baroni,zein} as implemented in the MBPP approach \cite{heid99},
which also provides direct access to the electron-phonon coupling
matrix elements. The procedure to extract the Eliashberg function is
outlined in Ref. \cite{heid10}. Dynamical matrices and corresponding EPC
matrix elements are obtained on a $16 \times 16 \times 16$ phonon mesh. These quantities
are then interpolated using standard Fourier techniques to the whole BZ,
and the Eliashberg functions are calculated by integration over the BZ
using the tetrahedron method on a $40 \times 40 \times 40$ mesh. SOC is consistently
taken into account in all calculations including lattice dynamical and EPC
properties. It is well known from previous work, that SOC is necessary
for a correct quantitative description of both the phonon anomalies and
EPC of undoped bulk lead \cite{heid10}.

Charge induction is simulated by adding an appropriate number of electrons
during the DFT self-consistency cycle, compensated by a homogeneous
background charge to retain overall charge neutrality.	Electronic
structure properties and the Eliashberg function are calculated for face centred cubic (fcc)
lead with the lattice constant $a=4.89$ {\AA} as obtained by optimization for
the undoped case.  For doping levels considered here, we found
that to a good approximation charge induction does not change the band
structure but merely results in a shift of $E_{F}$.  In a previous study,
the variation of the EPC was studied as function of the averaging energy
\cite{sklyadneva}. The present approach goes beyond this analysis by taking
into account explicitly the effect of charge induction on the screening
properties, which modifies both the phonon spectrum and the EPC matrix
elements.

Finally we point out that, since this DFT approach simulates the effect of the electric field by adding extra charge carriers to the system together with a uniform compensating counter-charge (Jellium model \cite{Pines}) is unable to describe inhomogeneous distributions caused by the screening of the electric field itself. A more complete approach has been developed in Ref. \onlinecite{Brumme}, and requires the self-consistent solution of the Poisson equation; however, this method is currently unable to compute the phonon spectrum of the gated material, making it unsuitable for the application of the proximity Eliashberg formalism.

\section{RESULTS AND DISCUSSION}
\begin{figure}
\begin{center}
\includegraphics[keepaspectratio, width=\columnwidth]{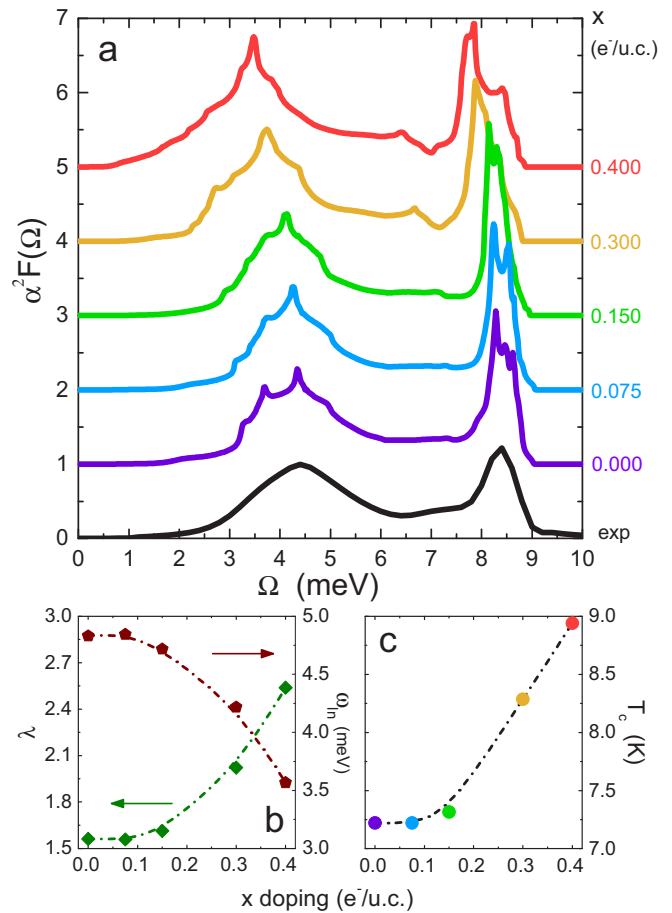}
\vspace{-5mm} \caption{(Color online)
 Panel a: calculated surface electron-phonon spectral function for five different value of charge doping (electrons/unitary cell) 0.00 (violet solid line), 0.075 (blue solid line), 0.15 (green solid line), 0.30 (orange solid line) and 0.40 (red solid line).  We also show the experimental electron-phonon spectral function determined via tunneling measurements (black solid line). All curves are shifted
 by a constant offset equal to one. Panel b: calculated values of electron-phonon coupling constants $\lambda$ (green diamonds and rhombus) and representative energies $\omega_{ln}$ (brown pentagons) versus charge doping. Panel c: calculated critical temperature versus charge doping for a system without proximity effect. All dash-dot lines acts as guides to the eye.
 }\label{Figure1}
\end{center}
\end{figure}

We start our calculations by fixing the input parameters for bulk lead according to the established literature. We set $T_{c,b}$ to its experimental value \cite{carbibastardo} $T_{c,b} = 7.22$ K. The undoped $\alpha^{2}_{b}F(\Omega)$ gives a corresponding electron-phonon coupling $\lambda_{b}=1.5596$. Assuming a cutoff energy $\omega_{c} = 60$ meV and a maximum energy $\omega_{max} = 70$ meV in the Eliashberg equations, we are thus able to determine the bulk Coulomb pseudopotential to be $\mu^{*}_{b} = 0.14164$ to obtain the exact experimental critical temperature $T_{c,b}$.

\begin{table*}
\begin{center}
\begin{tabular}{|c|c|c|c|c||c||c|}
  \hline
  $x (e^{-}/cell)$   & $\lambda$         & $\omega_{ln}$ $(meV)$  & $N(0)$ $states/(eV spin)$ & $\Delta E_{F}$ $(meV)$  & $\mu^{*}$ & $T_{c}$  $(K)$ \\
  0.000           & 1.5612            & 4.8431               & 0.25866               &   0.00             & 0.14164 & 7.2200 \\
  0.075           & 1.5582            & 4.8432               & 0.25754               & 108.42             & 0.14136 & 7.2197 \\
  0.150           & 1.6137            & 4.7176               & 0.25611               & 218.77             & 0.14116 & 7.3165 \\
  0.300           & 2.0237            & 4.2175               & 0.26770               & 435.07             & 0.14074 & 8.2862 \\
  0.400           & 2.5392            & 3.5668               & 0.27833               & 571.62             & 0.14048 & 8.9406 \\
  \hline
\end{tabular}
\caption{Calculated input parameters with DFT and  calculated critical temperature with Eliashberg theory without proximity effect.}\label{tab:exp}
\end{center}
\end{table*}

In Fig. \ref{Figure1}a we show the calculated electron-phonon spectral functions $\alpha^{2}F(\Omega)$ resulting at the increase of the doping level $x$. Specifically, we plot the curves corresponding to $x=0.000, 0.075, 0.150, 0.300 ,0.400$ e\apex{-}/unit cell.
We calculate the spectral functions until $x=0.4$ $e^{-}/$cell because for larger values of doping an instability emerges in the calculation processes.
We can see the phonon softening evidenced by a reduction of $\omega_{ln}$ with increasing doping level.
The increase of the carrier density gives rise to two competing effects: the value of $\omega_{ln}$ (i.e. the representative phonon energy) decreases while the value of electron-phonon coupling costant $\lambda$ increases (see Fig. \ref{Figure1}b). Since the critical temperature is an increasing function of both $\omega_{ln}$ and $\lambda$, in general this could result in either a net enhancement or suppression of $T_{c}$, depending on which of the two contributions is dominant. Consequently the ideal situation for obtaining largest critical temperature in an electric field doped material is to have a strong increase of $\lambda$ and $\omega_{ln}$ concurrently. In the case of lead the contribution from the increase of $\lambda$ is dominant over that from the reduction of $\omega_{ln}$, giving rise to a net increase of the superconducting critical temperature (as we report in Fig. \ref{Figure1}c). In addition, in Table \ref{tab:exp} we summarize all the input parameters of the proximity Eliashberg equations as obtained from the DFT calculations.

Having determined the response of the superconducting properties of a homogeneous lead film to a modulation of its carrier density, we can now consider the behavior of the more realistic junction between the perturbed surface layer and the unperturbed bulk. In order to do so, however, it is now mandatory to select a value for the thickness of the perturbed surface layer.
Close to $T_c$, the superfluid density is small \cite{HirschPRB2004} and the screening is dominated by unpaired electrons. Thus, a very rough approximation would be to set $d_{s}$ to the Thomas-Fermi screening length $d_{TF}$, which for lead can be estimated to be 0.15 nm \cite{Basavaiah}. However, we have recently shown \cite{PiattiNbN} that this assumption might not be satisfactory in the presence of the very large electric fields that build up in the electric double layer. Indeed, our experimental findings on niobium nitride indicated that the screening length increases for very large doping values \cite{PiattiNbN}. However, it is reasonable to assume the exact entity of this increase to be specific to each material. Thus, while the qualitative behavior can be expected to be general, the exact values of $d_{s}$ determined for niobium nitride cannot be directly applied to lead.

In order not to lose the generality of our approach, we calculate the behavior of our system for three different choices of the behavior of $d_{s}$. We start by expressing $d_{s} = d_{TF}[1 + m\Theta(x - x_{0})]$, where $m$ is a dimensionless parameter indicating how much $d_{s}$ expands beyond the Thomas-Fermi value for large doping levels, and $x_{0}$ is the specific doping value upon which this increase in $d_{s}$ takes place. By selecting $x_{0} = 0.2$, we allow the upper half of our doping values to go beyond the Thomas-Fermi approximation. We then perform proximity-coupled Eliashberg calculations for $m = 0,1,4$ and five different film thicknesses $d = 5, 10, 20, 30, 40$ nm, always assuming the junction area to be $A = 10^{-7}$ m\apex{2}. Note that the case $m = 0$ of course corresponds to the case where the material satisfies the Thomas-Fermi model for any value of doping: in this case, the model has no free parameters.

\begin{figure}[b]
\begin{center}
\includegraphics[keepaspectratio, width=0.9\columnwidth]{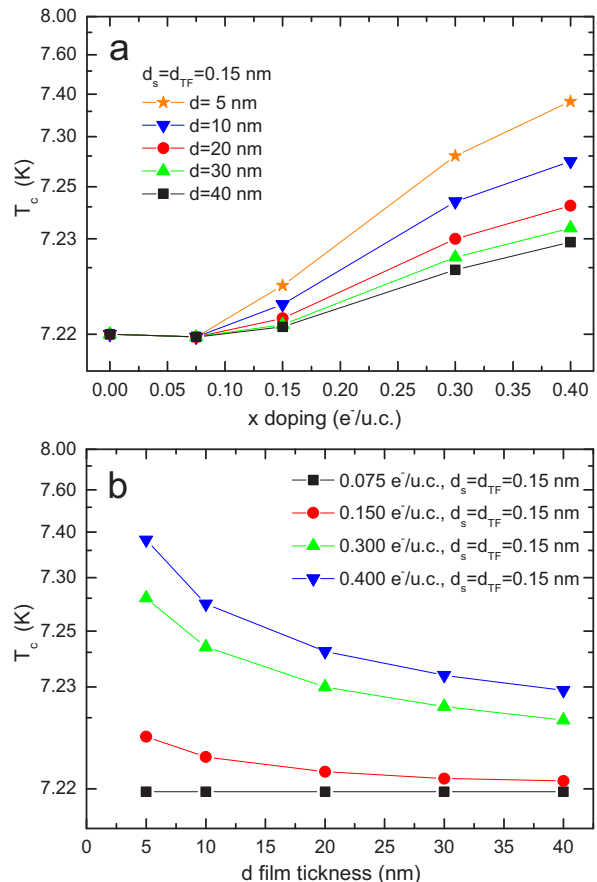}
\vspace{-5mm} \caption{(Color online)
 Panel a: calculated critical temperature versus charge doping for five different values of film thickness $d=5$ nm (orange stars), $d=10$ (blue down triangles), $d=20$ nm (red circles), $d=30$ nm (green up triangles) and $d=40$ nm (black squares) with surface layer thickness $d_{s}=d_{TF}$.
 Panel b: calculated critical temperature versus film thickness for four different charge doping (electrons/unitary cell): $0.075$ (black squares), $0.150$ (red circles),  $0.300$ (green up triangles), $0.400$ (blue down triangles) with $d_{s}=d_{TF}$. The two graphs are in semi-logarithmic scale ($log(T_{c} - 7.218)$).
 }\label{Figure2}
\end{center}
\end{figure}

In Fig. \ref{Figure2} we plot the evolution of $T_{c}$ upon increasing electron doping and assuming that the Thomas-Fermi model always holds ($m = 0$ and $d_{s} = d_{TF}$), for different values of film thickness.
The calculations show that the qualitative increase in $T_{c}$ with increasing doping level that we observed in the homogeneous case is retained also in proximized films of any thinckess (see Fig. \ref{Figure2}a). However, the presence of a coupling between surface and bulk induced by the proximity effect gives rise to a key difference with respect to the homogeneous case, namely, a strong dependence of $T_{c}$ on film thickness in the doped films. Indeed, the magnitude of the $T_{c}$ shift with respect to the homogeneous case is heavily suppressed already in films as thin as $5$ nm. This behavior is best seen in Fig. \ref{Figure2}b, where we plot the same data as a function of the total film thickness for all doping levels.
As we can see the increase of critical temperature drops dramatically with increasing film thickness. We have not calculated the critical temperature for monolayer films since the approximations of the model would no longer apply in this case: in particular the unperturbed electron-phonon spectral function would have been different from the bulk-like one we employed in our calculations \cite{Pratappone}.

\begin{figure}
\begin{center}
\includegraphics[keepaspectratio, width=0.9\columnwidth]{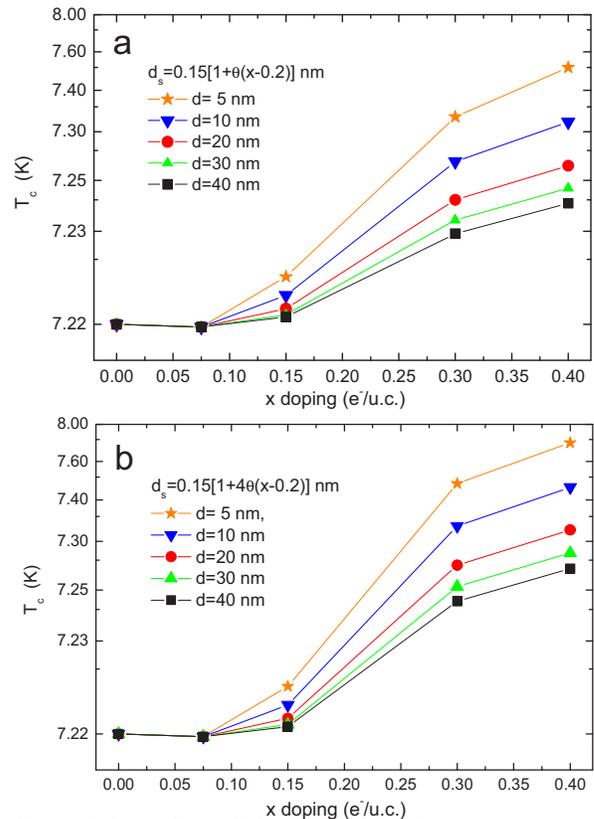}
\vspace{-5mm} \caption{(Color online)
 Calculated critical temperature versus charge doping for five different values of film thickness $d=5$ nm (orange stars), $d=10$ (blue down triangles), $d=20$ nm (red circles), $d=30$ nm (green up triangles) and $d=40$ nm (black squares) with surface layer thickness $d_{s}=d_{TF}[1+m\Theta(x-0.2)]$ (panel a $m=1$ and panel b $m=4$).  The two graphs are in semi-logarithmic scale ($log(T_{c} - 7.218)$).
 }\label{Figure3}
\end{center}
\end{figure}

We now consider the effect of the different degrees of confinement for the induced charge carriers at the surface of the films. We do so by allowing the perturbed surface layer to spread further in the depth of the film for large electron doping, i.e. by increasing the $m$ parameter in the definition of $d_s$.
In Fig. \ref{Figure3} we plot the evolution of $T_{c}$ with increasing electron doping and for different film thicknesses, in the two cases $m = 1$ ($d_s$ is allowed to expand up to $2d_{TF} = 0.3$ nm) and $m = 4$ ($d_s$ is allowed to expand up to $5d_{TF} = 0.75$ nm).
We can first observe how a different value of $d_s$ does not change the qualitative behavior of the films. The evolution of $T_c$ with increasing electron doping is still comparable to both the homogeneous case and the proximized films in the Thomas-Fermi limit. The suppression of the $T_c$ increase with increasing film thickness is also similar to the latter case. However, the magnitude of the $T_c$ shift \emph{for the same values of film thickness and doping level per unit cell} is clearly the more enhanced the larger the value of $d_s$. This is to be expected, as larger values of $d_s$ increase the fraction of the film that is perturbed by the application of the electric field and reduce the $T_{c}$ shift dampening operated by the proximity effect. In principle, for values of $m$ large enough (or film thickness $d$ small enough) one could reach the limit value $d_s \simeq d$ and recover the homogeneous case where the $T_c$ shift is maximum.

All the calculations we performed so far assumed that one could directly control the induced carrier density \emph{per unit volume}, $x$, in the surface layer, without an explicit upper limit. However, this is not an experimentally achievable goal in a field-effect device architecture. In this class of devices, the polarization of the gate electrode allows one to tune the electric field at the interface and thus the induced carrier density \emph{per unit surface}, $\Delta n_{2D}$, required to screen it, i.e. $\Delta n_{2D}=\int_{0}^{d_{s}} \Delta n_{3D}dz$ within our model is distributed within a layer of thickness $d_s$. In general, the determination of the exact depth profile of this distribution requires the self-consistent solution of the Poisson equation \cite{Brumme}; however, as a first approximation we can consider this distribution to be constant, obtaining an effective doping level per unit volume simply as $x =\Delta n_{2D}/d_s$. This procedure allows one to employ the same DFT-Eliashberg formalism we developed before in order to simulate a field effect experiment on a superconducting thin film.

\begin{figure*}
\begin{center}
\includegraphics[keepaspectratio, width=1.5\columnwidth]{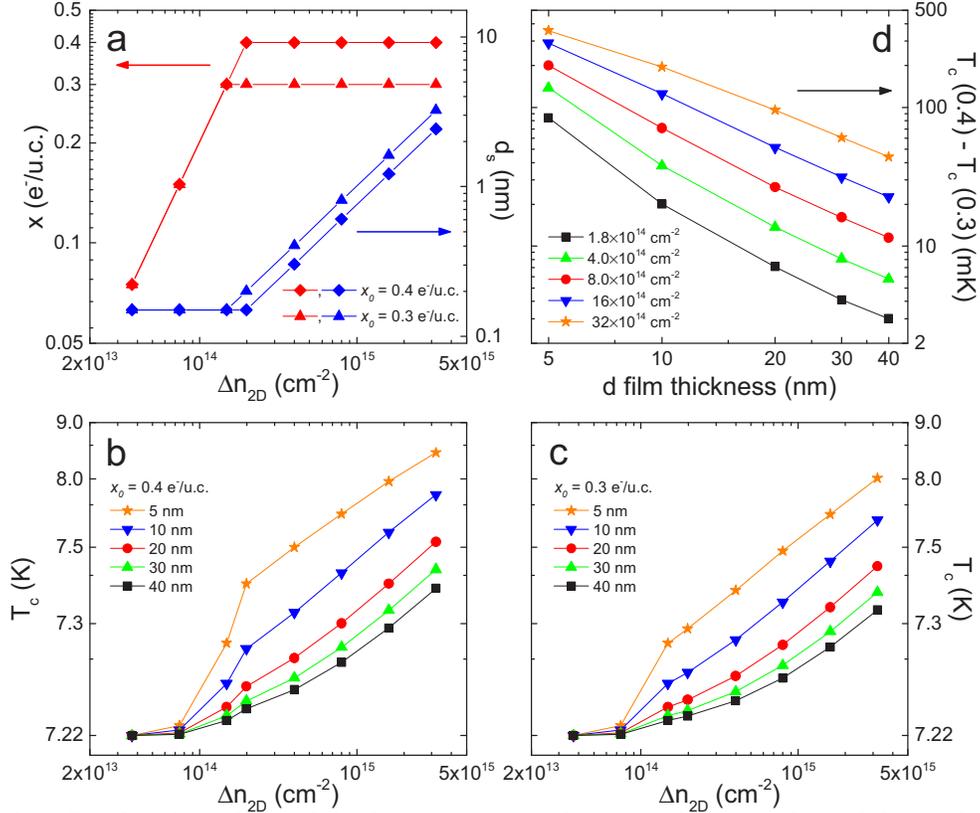}
\vspace{-5mm} \caption{(Color online)
 Panel a: dependence of the doping \emph{per unit volume} $x$ (red up triangles and diamonds) and surface layer thickness $d_{s}$ (blue up triangles and diamonds) on the induced carrier density \emph{per unit surface} $\Delta n_{2D}$, for two different values of the maximum doping level $x_{0} = 0.3$ and $x_{0} = 0.4$ e\apex{-}/unit cell.
 Panel b and panel c: $T_{c}$ versus induced carrier density \emph{per unit surface} $\Delta n_{2D}$ for five different film thicknesses ($d=5$ nm (orange stars), $d=10$ (blue down triangles), $d=20$ nm (red circles), $d=30$ nm (green up triangles) and $d=40$ nm (black squares) in the cases $x_{0} = 0.4$ and $0.3$ e\apex{-}/unit cell respectively. Panels b and c are in semi-logarithmic scale ($log(T_{c} - 7.2)$).
 Panel d: $T_{c}(x=0.4)-T_{c}(x=0.3)$ versus induced carrier density \emph{per unit surface} $\Delta n_{2D}$ for the five different film thicknesses.
 }\label{Figure4}
\end{center}
\end{figure*}

In addition, in the previous calculations we supposed that $d_s$ can only take on two values as a function of $x$, depending on the threshold value $x_0$. When we consider the field-effect architecture, however, the parameters $m$ and $x_0$ in the expression $d_{s} = d_{TF}[1 + m\Theta(x - x_{0})]$ are no longer independent as in the previous case. Moreover, according to our recent experimental findings on niobium nitride \cite{PiattiNbN}, $d_s$ is a monotonically increasing function of $\Delta n_{2D}$. We include this behavior in our calculations in the following way: Once the maximum doping level $x_{0}$ is selected, $m = m(\Delta n_{2D})$ is automatically determined by the requirement $d_{s}(\Delta n_{2D}) = \Delta n_{2D} \cdot x_0$ for any $x > x_{0}$. Fig. \ref{Figure4}a shows the resulting dependence of the doping \emph{per unit volume} $x$ and surface layer thickness $d_s$ on the induced carrier density \emph{per unit surface} $\Delta n_{2D}$, for two different values of the maximum doping level $x_{0} = 0.3$ and $x_{0} = 0.4$ e\apex{-}/unit cell. When $\Delta n_{2D}$ is small enough so that $x < x_{0}$, the Thomas-Fermi screening holds, $d_{s} = d_{TF}$ is constant and $x$ linearly increases with $\Delta n_{2D}$. As soon as $\Delta n_{2D}$ becomes large enough that $x = x_{0}$ is constant ($\Delta n_{2D}(x_0)$), the Thomas-Fermi screening is no longer valid and $d_{s}$ increases linearly with $\Delta n_{2D}$.

In Fig. \ref{Figure4}b and \ref{Figure4}c we plot the resulting modulation of $T_{c}$ for five different film thicknesses in the cases $x_{0} = 0.4$ and $0.3$ e\apex{-}/unit cell respectively. In both cases we can readily distinguish between two regimes of $\Delta n_{2D}$. When $\Delta n_{2D} \lesssim \Delta n_{2D}(x_{0})$, Thomas-Fermi screening holds and we reproduce the behavior we observed in Fig. \ref{Figure2}a. In this regime, the induced carrier density directly modulates $x$ and thus the electron-phonon spectral function $\alpha^2 F(\Omega)$. The $T_{c}$ modulation is thus a result of a direct modification of the material properties at the surface, with proximity effect simply operating a ``smoothing" the larger the value of the film thickness. On the other hand, when $\Delta n_{2D} > \Delta n_{2D}(x_{0})$, the surface properties ($\alpha^2 F(\Omega)$) are no longer modified by the extra charge carriers, and the further modulation of $T_{c}$ originates entirely from the proximity effect as determined by the increase in $d_{s}$. 

We can also compare the $T_c$ shifts for different maximum doping levels $x_0$. Fig. \ref{Figure4}d shows the difference between the $T_{c}$ corresponding to $x_{0} = 0.4$ and $0.3$ e\apex-/unit cell as a function of the total film thickness, for different values of $\Delta n_{2D}$. We can clearly see how $T_{c}$ is always larger for the films with larger $x_{0}$, for any value of film thickness, even if the associated values of $d_{s}$ are always smaller. This indicates that the maximum achievable value of $x_{0}$ is dominant with respect to the increase of $d_{s}$ to determine the final value of $T_{c}$, also in the doping regime $\Delta n_{2D} > \Delta n_{2D}(x_{0})$.

Of course, in a real sample we don't expect the transition between the two regimes to be so clear-cut, as the saturation of $x$ to $x_{0}$ would occur over a finite range of $\Delta n_{2D}$. In this intermediate region, the modulation of $\alpha^2 F(\Omega)$ and $d_{s}$ would both contribute in a comparable way to the final value of $T_{c}$ in the film. We stress, however, that in both regimes the proximity effect is fundamental in determining the $T_c$ of the gated film. We also note that the proximized Eliashberg equations are able to account for a non-uniform scaling of the $T_{c}$ shift for different values of film thickness, unlike the models that use approximated analytical equations for $T_{c}$.

\section{CONCLUSIONS}

In this work, we have developed a general method for the theoretical simulation of field-effect-doping in superconducting thin films of arbitrary thickness, and we have benchmarked it on lead as a standard strong-coupling superconductor. Our method relies on \emph{ab-initio} DFT calculations to compute how the increasing doping level $x$ per unit volume modifies the structural and electronic properties of the material (shift of Fermi level $\Delta E_F$, density of states $N(0)$, and electron-phonon spectral function $\alpha^2 F(\Omega)$). The Coulomb pseudopotential $\mu_*$ is determined by simple calculations from some of these parameters. The properties of the pristine thin film (critical temperature $T_{c}$, device area $A$ and total film thickness $d$) can be obtained either from the literature or experimentally from standard transport measurements. For doping values where the Thomas-Fermi theory of screening is satisfied, the perturbed surface layer thickness is constant ($d_s = d_{TF}$) and the theory has no free parameters.

Once all the input parameters are known, our method allows to compute the transition temperature $T_{c}$ for arbitrary values of film thickness $d$ and electron doping in the surface layer $x$ by solving the proximity-coupled Eliashberg equations in the surface layer and unperturbed bulk. On the other hand, if no reliable estimations of the surface layer thickness $d_{s}$ are available, our method allows one to determine $d_{s}(x)$ by reproducing the experimentally-measured $T_{c}(x)$. This allows to probe deviations from the standard Thomas-Fermi theory of screening in the presence of very large interface electric fields.

We also show how, even in the case where the Thomas-Fermi approximation breaks down and the doping level $x$ can no longer be increased, the transition temperature $T_{c}$ of a thin film can still be indirectly modulated by the electric field by changing the surface layer thickness $d_{s}$. For what concerns artificial enhancements of $T_c$ in superconducting thin films, we conclude that very thin films ($d \lesssim d_s$, in order to minimize the smoothing operated by the proximity effect) of a superconductive material characterized by a strong increase of the electron-phonon (boson) coupling upon changing its carrier density are required to optimize the effectiveness of the field-effect-device architecture.

Finally, our calculations indicate that sizable $T_c$ enhancements of the order of $\sim 0.5$ K should be achievable in thin films of a standard strong-coupling superconductor such as lead, for easily realizable thicknesses of $\sim 10$ nm and doping levels routinely induced via EDL gating in metallic systems. These features may open the possibility for superconducting switchable devices and electrostatically reconfigurable superconducting circuits above liquid helium temperature.

\begin{acknowledgments}
The work of G. A. U. was supported by the Competitiveness Program of NRNU MEPhI.
\end{acknowledgments}
%%%%%%%%%%%%%%%%%%%%%%%%%%


\begin{thebibliography}{99}
\small{
%%%%%%%%%%%%%%%%%%%%%%%%%%%%%%%%%%%%%%%%%%%%%%%%%%%%%%%%%%%%%%%%%%%%%%%%%%%%%%%%%%
\bibitem{FujimotoReview2013} T. Fujimoto and K. Awaga, \textit{Phys. Chem. Chem. Phys.} \textbf{15}, 8983 (2013)

\bibitem{UenoReview2014} K. Ueno, H. Shimotani, H. T. Yuan, J. T. Ye, M. Kawasaki, and Y. Iwasa, \textit{J. Phys. Soc. Jpn.} \textbf{83}, 032001 (2014)

\bibitem{GoldmanReview2014} A. M. Goldman, \textit{Annu. Rev. Mater. Res.} \textbf{44}, 45 (2014)

\bibitem{SaitoReview2016} Y. Saito, T. Nojima, and Y. Iwasa, \textit{Supercond. Sci. Technol.} \textbf{29}, 093001 (2016)

\bibitem{UenoNatureMater2008} K. Ueno, S. Nakamura, H. Shimotani, A. Ohtomo, N. Kimura, T. Nojima, H. Aoki, Y. Iwasa, and M. Kawasaki, \textit{Nat. Mater.} \textbf{7}, 855 (2008)

\bibitem{YeNatureMater2010} J. T. Ye, S. Inoue, K. Kobayashi, Y. Kasahara, H. T. Yuan, H. Shimotani, and Y. Iwasa, \textit{Nat. Mater.} \textbf{9}, 125 (2010)

\bibitem{SaitoScience2015} Y. Saito, Y. Kasahara, J. T. Ye, Y. Iwasa, and T. Nojima, Science \textbf{350}, 409 (2015)

\bibitem{UenoNatureNano2011} K. Ueno, S. Nakamura, H. Shimotani, H. T. Yuan, N. Kimura, T. Nojima, H. Aoki, Y. Iwasa, and M. Kawasaki, \textit{Nat. Nanotech.} \textbf{6}, 408 (2011)

\bibitem{YeScience2012} J. T. Ye, Y. J. Zhang, R. Akashi, M. S. Bahramy, R. Arita, and Y. Iwasa, \textit{Science} \textbf{338}, 1193 (2012)

\bibitem{JoNanoLett2015} S. Jo, D. Costanzo, H. Berger, and  A. F. Morpurgo, \textit{Nano Lett.} \textbf{15} 1197 (2015)

\bibitem{LuScience2015} J. M. Lu, O. Zheliuk, I. Leermakers, N. F. Q. Yuan, U. Zeitler, K. T. Law, and J. T. Ye, \textit{Science} \textbf{350}, 1353 (2015)

\bibitem{ShiSciRep2015} W. Shi, J. T. Ye, Y. Zhang, R. Suzuki, M. Yoshida, J. Miyazaki, N. Inoue, Y. Saito, and Y. Iwasa, \textit{Sci. Rep.} \textbf{5}, 12534 (2015)

\bibitem{YuNatNano2015} Y. Yu, F. Yang, X. F. Lu, Y. J. Yan, Y.-H. Cho, L. Ma, X. Niu, S. Kim, Y.-W. Son, D. Feng, S. Li, S.-W. Cheong, X. H. Chen, and Y. Zhang, \textit{Nat. Nanotechnol.} \textbf{10}, 270 (2015)

\bibitem{CostanzoNatNano2016} D. Costanzo, S. Jo, H. Berger, and A. F. Morpurgo, \textit{Nat. Nanotechnol.} \textbf{11}, 399 (2016)

\bibitem{SaitoNatPhys2016} Y. Saito, Y. Nakamura, M. S. Bahramy, Y. Kohama, J. T. Ye, Y. Kasahara, Y. Nakagawa, M. Onga, M. Tokunaga, T. Nojima, Y. Yanase, and Y. Iwasa, \textit{Nat. Phys.} \textbf{12}, 144 (2016)

\bibitem{bollinger11} A. T. Bollinger, G. Dubuis, J. Yoon, D. Pavuna, J. Misewich, and I. Bo{\v{z}}ovi{\'c}, \textit{Nature} \textbf{472}, 458 (2011)

\bibitem{LengPRL2011} X. Leng, J. Garcia-Barriocanal, S. Bose, Y. Lee, and A. M. Goldman, \textit{Phys. Rev. Lett.} \textbf{107}, 027001 (2011)

\bibitem{LengPRL2012} X. Leng, J. Garcia-Barriocanal, B. Yang, Y. Lee, J. Kinney, and A. M. Goldman, \textit{Phys. Rev. Lett.} \textbf{108}, 067004 (2012)

\bibitem{MaruyamaAPL2015} S. Maruyama, J. Shin, X. Zhang, R. Suchoski, S. Yasui, K. Jin, R. L. Greene, and I. Takeuchi, \textit{Appl. Phys. Lett.} \textbf{107}, 142602 (2015)

\bibitem{JinSciRep2016} K. Jin, W. Hu, B. Zhu, J. Yuan, Y. Sun, T. Xiang, M. S. Fuhrer, I. Takeuchi, and R. L. Greene, \textit{Sci. Rep.} \textbf{6}, 26642 (2016)

\bibitem{FeteAPL2016} A. F{\^e}te, L. Rossi, A. Augieri, and C. Senatore, \textit{Appl. Phys. Lett.} \textbf{109}, 192601 (2016)

\bibitem{BurlachkovPRB1993} L. Burlachkov, I. B. Khalfin, and B. Ya. Shapiro, \textit{Phys. Rev. B} \textbf{48}, 1156 (1993)

\bibitem{GhinovkerPRB1995} M. Ghinovker, V. B. Sandomirsky, and B. Ya. Shapiro, \textit{Phys. Rev. B} \textbf{51}, 8404 (1995)

\bibitem{walter16} J. Walter, H. Wang, B. Luo, C. D. Frisbie, and C. Leighton, \textit{ACS Nano} \textbf{10}, 7799 (2016)

\bibitem{libro} P. Lipavsky, J. Kolacek, K. Morawetz, E. H. Brandt, and T.-J. Yang, Bernoulli Potential in
Superconductors How the Electrostatic Field Helps to Understand Superconductivity, \textit{Lect. Notes
Phys.} \textbf{733}, Springer, Berlin Heidelberg (2008)

\bibitem{LiNature2016} L. J. Li, E. C. T. O'Farrell, K. P. Loh, G. Eda, B. {\"O}zyilmaz, and A. H. Castro Neto, \textit{Nature} \textbf{529}, 185 (2016)

\bibitem{YoshidaAPL2016} M. Yoshida, J. T. Ye, T. Nishizaki, N. Kobayashi, and Y. Iwasa, \textit{Appl. Phys. Lett.} \textbf{108}, 202602 (2016)

\bibitem{XiPRL2016} X. X. Xi, H. Berger, L. Forr{\'o}, J. Shan, and K. F. Mak, \textit{Phys. Rev. Lett.} \textbf{117}, 106801 (2016)

\bibitem{ShiogaiNaturePhys2015} J. Shiogai, Y. Ito, T. Mitsuhashi, T. Nojima, and A. Tsukazaki, \textit{Nat. Phys}. \textbf{12}, 42 (2016)

\bibitem{LeiPRL2016} B. Lei, J. H. Cui, Z. J. Xiang, C. Shang, N. Z. Wang, G. J. Ye, X. G. Luo, T. Wu, Z. Sun, and X. H. Chen, \textit{Phys. Rev. Lett.} \textbf{116}, 077002 (2016)

\bibitem{HanzawaPNAS2016} K. Hanzawa, H. Sato, H. Hiramatsu, T. Kamiya, and H. Hosono, \textit{Proc. Natl. Acad. Sci. USA} \textbf{113}, 3986 (2016)

\bibitem{ChoiAPL2014} J. Choi, R. Pradheesh, H. Kim, H. Im, Y. Chong, and D. H. Chae, \textit{Appl. Phys. Lett} \textbf{105}, 012601 (2014)

\bibitem{PiattiJSNM2016} E. Piatti, A. Sola, D. Daghero, G. A. Ummarino, F. Laviano, J. R. Nair, C. Gerbaldi, R. Cristiano, A. Casaburi, and R. S. Gonnelli, \textit{J. Supercond. Novel Magn.} \textbf{29}, 587 (2016)

\bibitem{PiattiNbN} E. Piatti, D. Daghero, G. A. Ummarino, F. Laviano, J. R. Nair, R. Cristiano, A. Casaburi, C. Portesi, A. Sola, and R. S. Gonnelli, \textit{Phys. Rev. B} \textbf{95}, 140501 (2017)

\bibitem{carbibastardo} J. P. Carbotte, \textit{Rev. Mod. Phys.} \textbf{62}, 1027 (1990)

\bibitem{ummarinorev} G. A. Ummarino, Eliashberg Theory. In: \textit{Emergent Phenomena in Correlated Matter}, edited by E. Pavarini, E. Koch, and U. Schollw\"{o}ck, Forschungszentrum J\"{u}lich GmbH and Institute for Advanced Simulations, pp.13.1-13.36 (2013) ISBN 978-3-89336-884-6

\bibitem{Mc} W.L. McMillan, \textit{Phys. Rev.} \textbf{175}, 537, (1968)

\bibitem{Carbi1} E. Schachinger and J. P. Carbotte, \textit{J. Low Temp. Phys.} \textbf{54}, 129 (1984)

\bibitem{Carbi2} H. G. Zarate and J. P. Carbotte, \textit{Phys. Rev. B} \textbf{35}, 3256, (1987)

\bibitem{Carbi3} H. G. Zarate and J. P. Carbotte, \textit{Physica B+ C} \textbf{135}, 203 (1985)

\bibitem{kresin} V. Z. Kresin, H. Morawitz, and S. A. Wolf, \textit{Mechanisms of Conventional and High Tc Superconductivity}, Oxford University Press (1999)

\bibitem{UmmaC60} G. A. Ummarino and R. S. Gonnelli, \textit{Phys. Rev. B} \textbf{66}, 104514 (2002)

\bibitem{Morel} P. Morel and P. W. Anderson, \textit{Phys. Rev.} \textbf{125}, 1263 (1962)

\bibitem{pastore} G. Grosso and G. Pastori Parravicini, \textit{Solid state Physics}, Academic Press (2014), ISBN 9780123850300

\bibitem{Grimvall} G. Grimvall, \textit{The Electron-phonon Interaction in Metals}, North Holland, Amsterdam (1981); G. A. Ummarino, \textit{Physica C} \textbf{423}, 96102 (2005)

\bibitem{Thomas} F. Stern, \textit{Phys. Rev. Lett.} \textbf{18}, 546 (1967); E. Canel, M. P. Matthews, and R. K. P. Zia, \textit{Phys. Kondens. Mater.} \textbf{15}, 191 (1972); J. Lee and H. N. Spector, \textit{J. Appl. Phys.} \textbf{54}, 6989 (1983)

\bibitem{meyer} B. Meyer, C. Els\"{a}sser, and M. F\"{a}hnle, FORTRAN90 Program for Mixed-Basis Pseudopotential Calculations for Crystals, Max-Planck-Institut f\"{u}r Metallforschung, Stuttgart (unpublished)

\bibitem{vanderbilt} D. Vanderbilt, \textit{Phys. Rev. B} \textbf{32}, 8412 (1985)

\bibitem{hedin} L. Hedin and B. I. Lundqvist, \textit{J. Phys. C} \textbf{4}, 2064 (1971)

\bibitem{heid99} R. Heid and K. P. Bohnen, \textit{Phys. Rev. B} \textbf{60}, R3709 (1999)

\bibitem{heid10} R. Heid, K.-P. Bohnen, I. Yu. Sklyadneva, and E. V. Chulkov, \textit{Phys. Rev. B} \textbf{81}, 174527 (2010)

\bibitem{baroni} S. Baroni, S. de Gironcoli, A. Dal Corso, and P. Giannozzi, \textit{Rev. Mod. Phys.} \textbf{73}, 515 (2001)

\bibitem{zein} N. E. Zein, \textit{Fiz. Tverd. Tela (Leningrad)} \textbf{26}, 3028 (1984); N. E. Zein, \textit{Sov. Phys. Solid State} \textbf{26}, 1825 (1984)

\bibitem{sklyadneva} I. Yu. Sklyadneva, R. Heid, P. M. Echenique, K.-B. Bohnen, and E. V. Chulkov, \textit{Phys Rev. B} \textbf{85}, 155115 (2012)

\bibitem{Pines} D. Pines and P. Nozieres, \textit{The theory of quantum liquids}, Benjamin, New York (1966)

\bibitem{Basavaiah} S. Basavaiah, J. M. Eldridge, and J. Matisoo, \textit{J. of Appl. Phys.} \textbf{45}, 457 (1974)

\bibitem{Pratappone} S. Bose, C. Galande, S. P. Chockalingam, R. Banerjee, P. Raychaudhuri, and P. Ayyub, \textit{J. Phys.: Condens. Matter} \textbf{21}, 205702 (2009)

\bibitem{Brumme} T. Brumme, M. Calandra, and F. Mauri, \textit{Phys Rev. B} \textbf{91}, 155436 (2012)

\bibitem{HirschPRB2004} J. E. Hirsch, \textit{Phys. Rev. B} \textbf{70}, 226504 (2004)
%%%%%%%%%%%%%%%%%%%%%%%%%%%%%%%%%%%%%%%%%%%%%%%%%%%%%%%%%%%%%%%%%%%%%%%%%%%%%%%%%%%%%%%%%%%%
%
%%%%%%%%%%%%%%%%%%%%%%%%%%%%%%%%%%%%%%%%

}\end{thebibliography}
\end{document}